\documentclass[pre,twocolumn,a4paper,superscriptaddress,floatfix,10pt,nofootinbib,longbibliography]{revtex4-2}
\usepackage{graphicx}
\usepackage{dcolumn}
\usepackage{bm}
\usepackage{color}
\usepackage{amsmath}
\usepackage{amssymb}
\usepackage{hyperref}
\usepackage{algorithm}
\usepackage[end]{algpseudocode}

\newcommand{\ci}{\mathrm{c}}

\newcommand{\beq}{\begin{equation}}
\newcommand{\eeq}{\end{equation}}
\newcommand{\ba}{\begin{array}}
\newcommand{\ea}{\end{array}}
\newcommand{\bea}{\begin{eqnarray}}
\newcommand{\eea}{\end{eqnarray}}

\begin{document}

\title{Scalable neural networks for the efficient learning of disordered quantum systems}

\author{N. Saraceni}
\affiliation{School of Science and Technology, Physics Division, Universit{\`a}  di Camerino, 62032 Camerino (MC), Italy}
\author{S. Cantori}
\affiliation{School of Science and Technology, Physics Division, Universit{\`a}  di Camerino, 62032 Camerino (MC), Italy}
\author{S. Pilati}
\affiliation{School of Science and Technology, Physics Division, Universit{\`a}  di Camerino, 62032 Camerino (MC), Italy}

\begin{abstract}
Supervised machine learning is emerging as a powerful computational tool to predict the properties of complex quantum systems at a limited computational cost.
In this article, we quantify how accurately deep neural networks can learn the properties of disordered quantum systems as a function of the system size.
We implement a scalable convolutional network that can address arbitrary system sizes. This network is compared with a recently introduced extensive convolutional architecture [K. Mills \emph{et al.}, Chem. Sci. {\bf 10}, 4129 (2019)] and with conventional dense networks with all-to-all connectivity.
The networks are trained to predict the exact ground-state energies of various disordered systems, namely a continuous-space single-particle Hamiltonian for cold-atoms in speckle disorder, and different setups of a quantum Ising chain with random couplings, including one with only short-range interactions and one augmented with a long-range term.
In all testbeds we consider, the scalable network retains high accuracy as the system size increases.
Furthermore, we demonstrate that the network scalability enables a transfer-learning protocol, whereby a pre-training performed on small systems drastically accelerates the learning of large-system properties, allowing reaching high accuracy with small training sets.
In fact, with the scalable network one can even extrapolate to sizes larger than those included in the training set, accurately reproducing the results of  state-of-the-art quantum Monte Carlo simulations.
\end{abstract}

\maketitle

\section{Introduction}
\label{secintro}
Supervised machine-learning algorithms allow one to train sophisticated statistical models to associate the different instances of a quantum system to the corresponding physical properties~\cite{dunjko2018machine,carleo2019machine,carrasquilla2020machine}. 
%
%
%
%
These algorithms have already been used to boost some of the most important computational tasks in  quantum chemistry and in material science, including:  molecular dynamics simulations~\cite{blank1995neural,behler2007generalized,behler2011neural,bartok2017machine,zhang2018end},  electronic-structure calculations based on density-functional theory~\cite{snyder2012finding,li2016understanding,brockherde2017bypassing,ryczko2019deep,moreno2019deep}, and molecular-property predictions  from structural information~\cite{hansen2013assessment,schutt2014represent,hansen2015machine}.
They have also been used  in drug-design research to predict binding affinities of protein-molecule complexes~\cite{ballester2010machine,khamis2015machine,jimenez2018k}.
Several statistical models have been adopted, including kernel-ridge regression, support vector machines, random forest, and artificial neural networks.
Like in many other fields of science and engineering, deep neural networks are emerging as the most promising candidates. This is mostly due to their ability to automatically extract the relevant features out of many system descriptors~\cite{mills2017deep} --- thus avoiding hand-crafted features --- and to systematically improve in accuracy as the amount of training data increases~\cite{pilati2019supervised}.
However, for small training sets the deep networks are plagued by the overfitting problem, meaning that they fail to accurately generalize to previously unseen instances. This is a critical problem, given that producing copious training sets for large quantum systems is computationally unfeasible, unless one accepts crude approximations.

In this article, we investigate the scalability of various neural networks in the supervised learning of  quantum systems. For scalability, we mean their ability to address arbitrary system sizes -- without changing the network structure, namely, the number of neurons and of connecting weights -- while maintaining satisfactory performance as the system size increases.
One of our main goals is to develop statistical models that provide accurate predictions for large quantum systems, even when the training set is sparse.
In previous studies performed in the field of molecular dynamics, approximately scalable models have been implemented by computing potential energies  as the sum of single-atom contributions~\cite{behler2016perspective}. These contributions take into account the atomic environment, but  only up to a finite cut-off distance.
This approach could lead to uncontrolled approximations in the presence of long-range interactions or correlations~\cite{bartok2017machine}.
In quantum chemistry, a limited form of scalability has been achieved by representing instances of different sizes using the same number of descriptors~\cite{rupp2012fast,hansen2015machine,huo2017unified,stuke2019chemical,jungsize} (usually extending with zeros the descriptor vectors of the small instances).
%
%
%
These descriptors are fed to intrinsically non-scalable models (i.e., models accepting a fixed-number of input descriptors) as, e.g., kernel ridge regression. 
This approach becomes impractical for large systems, and it does do not allow extrapolating beyond the sizes included in the training set.
Very recently, a scalable neural network designed to learn the extensive properties of solids has been introduced~\cite{mills2019extensive}. It is based on partitioning the systems into partially overlapping tiles and computing the (extensive) target value as the sum of single-tile contributions. 
The overlap regions ought to take into account boundary effects. 
The possible drawback of this model is that it appears not to be suitable for intensive properties. 
%
The statistical models mentioned above have been trained and tested against approximate predictions based on density functional theory and, for small chemical systems, on the coupled-cluster method. Therefore, it has not been verified whether they can actually learn the exact properties of large quantum many-body systems, or merely the simplification implied by the adopted approximate theories.
In this article, we introduce a novel architecture based on a deep convolutional neural network complemented by a global pooling layer. This layer allows the network addressing arbitrary system sizes, without necessarily retraining on each size.
Furthermore, it allows us to adopt a transfer learning technique~\cite{caruana1997multitask}, whereby the learning of large systems is accelerated by a pre-training performed on smaller sizes~\cite{zen2020transfer}. As we demonstrate, scalability also allows us to perform extrapolations to sizes larger than the ones included in the training set~\cite{efthymiou2019super,mills2019extensive}.
This novel scalable model is compared to the extensive architecture of Ref.~\cite{mills2019extensive}, and also to (non-scalable) dense networks with all-to-all interlayer connectivity.
To rigorously quantify the performances of these networks, we benchmark their predictions against the exact ground-state energies of disordered quantum Hamiltonians. 
In fact, synthetic disordered systems have emerged as suitable stringent testbeds for deep neural networks~\cite{mills2017deep,pilati2019supervised,ryczko2019deep} in the quantum-physics domain.
The first testbed system we consider is a one-dimensional continuous-space Hamiltonian for a single particle in a disordered potential. 
This model describes early cold-atom experiments on Anderson localization~\cite{roati2008anderson,billy2008direct}. 
In this case, the  ground-state energy is not extensive, meaning that it does not increase when the size of the optical field increases.
The second testbed is an (extensive) quantum Ising chain with disordered couplings. We consider a setup with only nearest-neighbor interactions, and also one augmented with far-neighbor interactions and one with frustrated couplings. The ferromagnetic quantum critical point is also addressed.
The ground-state energies are exactly computed via the Jordan-Wigner transformation, via exact diagonalization, and via unbiased quantum Monte Carlo simulations, depending on the specific setup.

Our findings indicate that the scalable model with the global pooling layer retains high accuracy for increasing system size, both for the single-particle Hamiltonian and for the quantum Ising chain. The extensive network performs well only in the latter testbed.
Transfer learning drastically accelerates the learning of large systems, allowing reaching high accuracy with training sets two orders of magnitude smaller than the ones needed without pre-training. 
Remarkably, the global pooling network is able to accurately extrapolate the ground-state energies of large Ising chains, including sizes that can be addressed only via computationally expensive quantum Monte Carlo simulations.

The rest of the article is organized as follows: the testbed systems, namely the continuous-space Hamiltonian and the different setups of the quantum Ising chain, are described in Section~\ref{secmodels}.
The networks and the training algorithm are described in Section~\ref{secnetworks}.
Our results on the supervised learning are reported in Section~\ref{secresults}.
Section~\ref{secconclusions} summarizes the main findings, with our conclusions and some future perspectives.

\begin{figure}[h]
\begin{center}
\includegraphics[width=1.0\columnwidth]{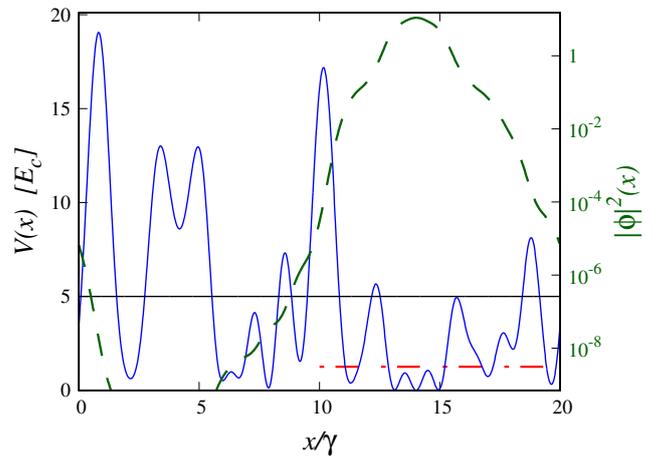}
\caption{(color online). 
Profile of an instance of an optical speckle field $V(x)$ (blue curve) as a function of the spatial coordinate $x/\gamma$. The system size is $L=20\gamma$ where $\gamma$ is the disorder correlations length. The (black) horizontal line represents the average intensity $V_0=\left<V(x)\right>$, while the (red) dot-dashed  segment indicates the ground-state energy $E$. Energies are expressed in units of the disorder correlation energy $E_c$.
The dashed (dark green) curve represents the squared modulus of the ground-state wave function $\phi(x)$ with $\ell_2$ normalization. The corresponding (logarithmic) scale is indicated on the right vertical axis.
}
\label{fig1}
\end{center}
\end{figure}
%

\begin{figure}[h]
\begin{center}
\includegraphics[width=0.6\columnwidth]{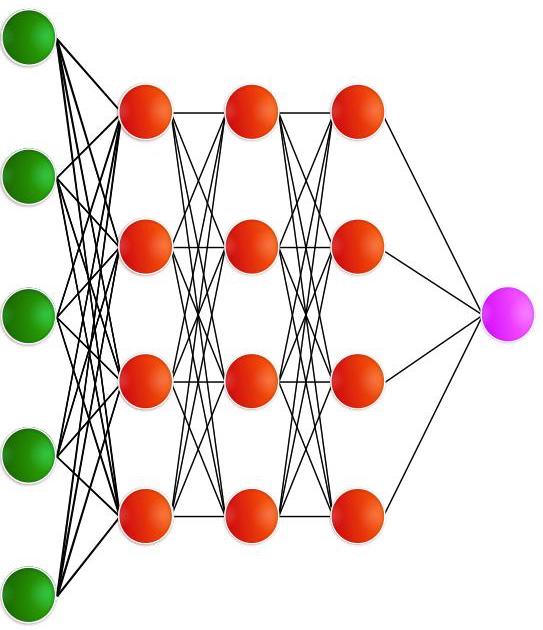}
\caption{(color online). 
Representation of the connectivity structure of a dense neural network. The (green) leftmost layer represents the $N_d$ input neurons. The three intermediate (red) layers include $N_n$ hidden neurons each. The output neuron on the right is associated to the target value. All neurons are connected to all neurons in the adjacent layers.
}
\label{fig2}
\end{center}
\end{figure}
%

\begin{figure}[h]
\begin{center}
\includegraphics[width=1.0\columnwidth]{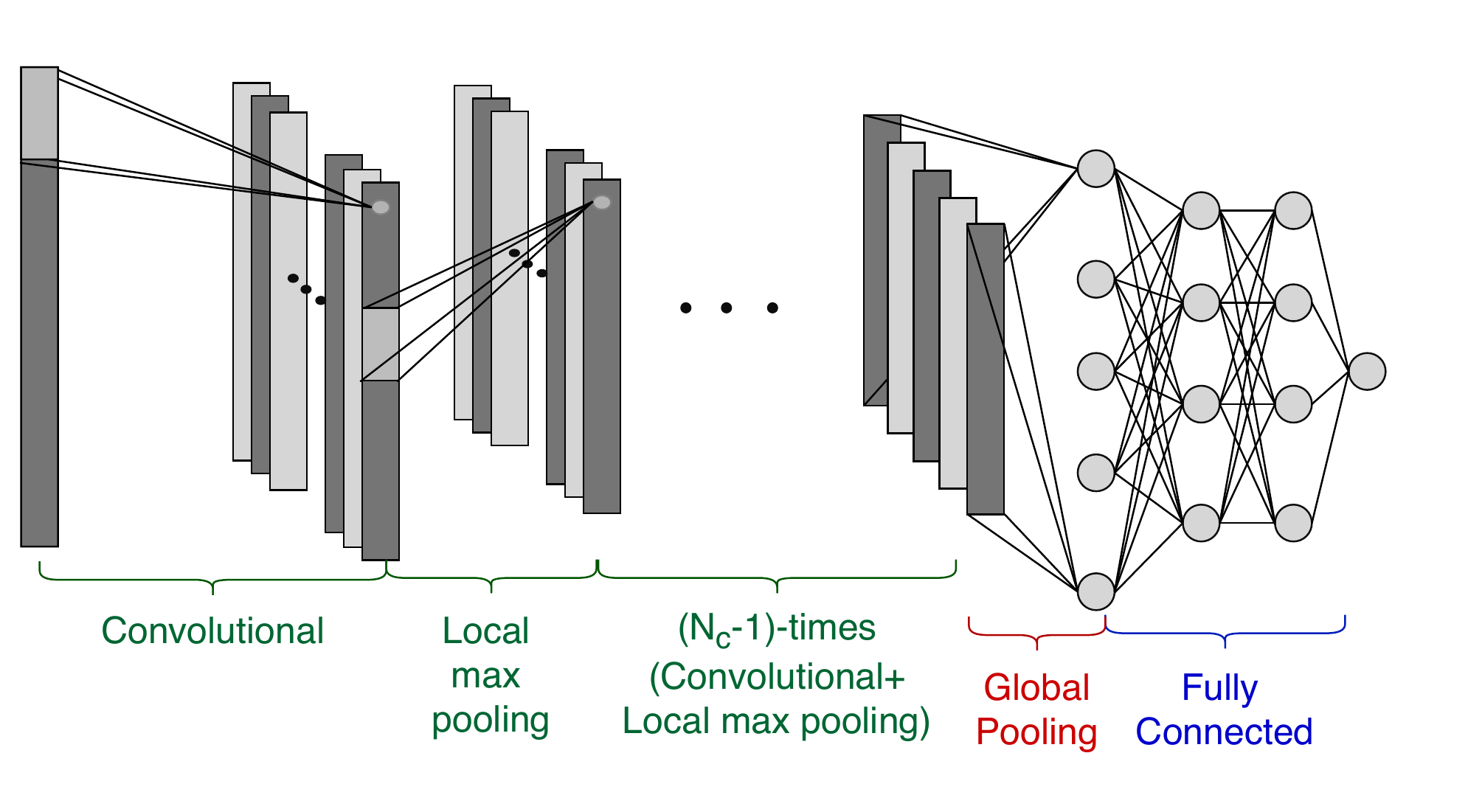}
\caption{(color online). 
Schematic representation of the scalable convolutional network with the global pooling layer.
The input layer (on the left) is followed by $N_c$ convolutional layers with $N_f$ filters, which create filtered maps of the input. 
The maps are  organized in depth. Local pooling layers are inserted between the convolutional layers, reducing the number of neurons. 
The convolutional layers are connected to the dense layers (on the right) via the global pooling layer. 
}
\label{fig3}
\end{center}
\end{figure}
%

\begin{figure}[h]
\begin{center}
\includegraphics[width=0.85\columnwidth]{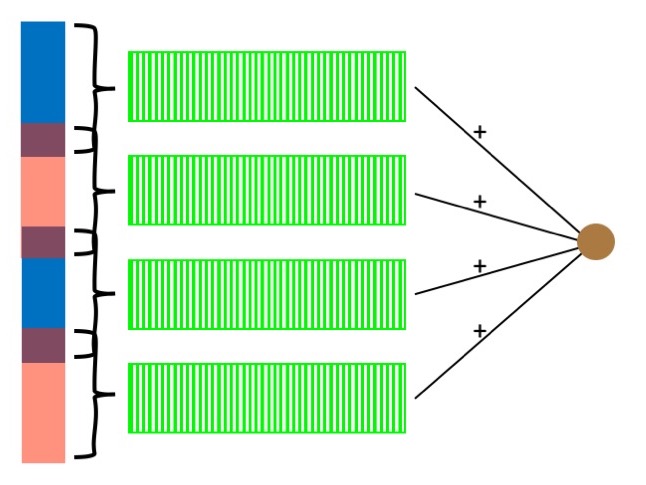}
\caption{(color online). 
Caricature of the extensive convolutional network, inspired by the architecture introduced in Ref.~\cite{mills2019extensive}.
The input system is divided into partially overlapping tiles, represented by alternating colors. The descriptors of each tile are fed into parallel identical  networks represented by the (green) rectangles with stripes. The outputs of these parallel networks are summed to obtain the target value, which is the total ground-state energy.
}
\label{fig4}
\end{center}
\end{figure}
%

\section{Testbed models}
\label{secmodels}
The first testbed system we consider is a one-dimensional continuous-space Hamiltonian for a single quantum particle. This Hamiltonian is defined as:
\begin{equation}
\hat{H}=-\frac{\hbar^2}{2m} \frac{\mathrm{d}^2}{\mathrm{d}x^2} + V(x),
\label{H}
\end{equation}
where $\hbar$ is the reduced Planck constant and $m$ is the particle mass. Periodic boundary conditions are adopted.
$V(x)$ is a random external potential designed to describe the effect of optical speckle patterns on alkali atoms.
The Hamiltonian~\eqref{H} describes the early experiments on the Anderson localization phenomenon performed with ultracold atomic gases~\cite{roati2008anderson,billy2008direct}.
Experimentally, optical speckle patterns are generated by shining coherent light through rough semitransparent surfaces. The transmitted light is then focused onto the atomic cloud.
%
%
In computer simulations, the corresponding potential can be generated using the numerical algorithm described in Refs.~\cite{huntley1989speckle,PhysRevA.73.013606}.
The generated potential satisfies periodic boundary conditions.
The potential is defined on a discrete spatial grid: $x_g=g\delta x$, where $\delta x=L/N_g$, $L$ is the system size, and the integer $g=0,1,\dots, N_g-1$. The number of grid points $N_g$  has to be large enough, as discussed below. 
For the blue-detuned optical fields considered in this article, the probability distribution of the local potential intensities $V_{\mathrm{loc}}=V(x)$, for any $x\in[0,L]$, is $P(V_{\mathrm{loc}})=\exp(-V_{\mathrm{loc}}/V_0)/V_0$ for $V_{\mathrm{loc}}\geqslant 0$, and $P(V_{\mathrm{loc}})=0$ otherwise. 
The parameter $V_0 \geqslant 0$ fixes the average intensity,  $\langle V(x) \rangle=V_0$. 
Different instances of the optical speckle field can be generated using in the numerical algorithm different pseudo-random numbers.
The average intensity $V_0$ also coincides with the the standard deviation of the speckle field: $\sqrt{\langle V(x)^2 \rangle-V_0^2}=V_0$.
Therefore, $V_0$ is the only parameter that determines the amount of disorder in the system.
We normalize the optical speckle field so that its spatial average over the finite system size $L$ is equal to $V_0$, for each individual instance, eliminating small fluctuations due to finite size effects.
The local intensities at two positions ${x}^\prime+{x}$ and ${x}^\prime$ are statistical correlated. The decay of these correlations is characterized by the following autocorrelation function:
\begin{equation}
\Gamma(x)=\frac{\langle V({ x}^\prime)V({x}^\prime+{x})\rangle}{V_0^2}-1 = \frac{\sin^2\left(\pi x/\gamma\right)}{\left(\pi  x/\gamma\right)^2}.
\label{speckle6}
\end{equation}
The parameter $\gamma$ characterizes length-scale of the spatial correlations.
%
The ground-state wave-function $\phi(x)$ of the Hamiltonian~\eqref{H} and the corresponding energy $E$ are determined via exact numerical diagonalization of the matrix obtained from a finite-difference approach. Specifically, the second derivative is represented via an $11$-point formula. 
For all system sizes, we set the number of grid points $N_g$ so that $\delta x = 0.0195 \gamma$. This is sufficiently small to essentially eliminate any discretization error.
We consider different instances of the optical speckle field, fixing the average intensity at the moderately large value $V_0=5E_{\ci}$, where $E_{\ci}= \hbar^2/(2m \gamma^2)$ is the correlation energy.
This intensity is sufficiently strong to observe the Anderson localization phenomenon in the finite sizes considered in this article~\cite{PhysRevA.100.013603,falco2010density}.
As a consequence of Anderson localization, the wave function $\phi(x)$ has non-negligible values only in a small region of space. 
Away from this core region, $\phi(x)$ displays an approximately exponential decay.
This effect is visualized in Fig.~\ref{fig1} for an instance of the speckle field of size $L=20\gamma$. 
The profile of the speckle potential $V(x)$ is also shown, together with corresponding ground-state energy $E$. 
This energy level randomly fluctuates for different instances of the speckle field. In Section~\ref{secresults}, various networks are trained to predict the ground-state energies of new speckle-field instances.
%

The second testbed system we consider is a quantum Ising chain. In general, this  model is  defined as:
\begin{equation}
\hat{H}=-\sum_{j=1}^{N_s} J_j{\sigma}^{z}_{j} {\sigma}^{z}_{j+1}- \sum_{j=1}^{N_s} J^{\prime}_j{\sigma}^{z}_{j} {\sigma}^{z}_{j+d} -\Gamma \sum_{j=1}^{N_s} {\sigma}^{x}_{j}.
\label{H2}
\end{equation}
$\sigma^x_j$ and $\sigma^z_j$ are conventional Pauli matrices at the lattice sites $j=1,\dots,N_s$. $N_s$ is the number of spins, and we consider again periodic boundary conditions. 
The couplings $J_j$ fix the strength of the  interactions between the nearest-neighbor spins $j$ and $j+1$.
The couplings $J^{\prime}_j$ fix the one between the spins $j$ and $j+d$. The integer $d>1$ fixes the range of this interaction term, as specified below.
$\Gamma$ is the intensity of the (uniform) transverse magnetic field. 
We consider various setups of the Hamiltonian~\eqref{H2}. In the first setup, beyond nearest-neighbor couplings are set to $J^{\prime}_j =0$. 
The nearest-neighbor couplings $J_j$ are sampled from the uniform probability distribution $\mathcal{P}(J)=\theta(J+1)\theta(1-J)$, where $\theta(x)$ is the unit step function: $\theta(x)=1$ for $x>0$ and $\theta(x)=0$ otherwise. %
The transverse field intensity is set at $\Gamma=0.5$.
%
We also consider a second setup with nonnegative couplings sampled from the distribution $\mathcal{P}_{>}(J)=\theta(J)\theta(1-J)$, with the transverse-field tuned at the critical point  $\Gamma \cong 0.36792$ of the ferromagnetic quantum phase transition~\cite{shankar1987nearest}.
The target value of the supervised learning procedure is, again, the ground-state energy $E$. In some cases, we consider the energy per spin $E/N_s$, as specified in the Appendix~\ref{Appendix}.
The ground-state energy can be exactly  computed at a modest computational cost by performing a Jordan-Wigner transformation to a free fermion model~\cite{pfeuty1970one}. For this computation, we follow the numerical algorithm of Ref.~\cite{young1996numerical}.
The third setup we consider for the Hamiltonian~\eqref{H2} includes both nearest-neighbor and next-nearest neighbor couplings. The latter corresponds to the range $d=2$. Both couplings are sampled from the (nonnegative) uniform distribution  $\mathcal{P}_{>}(J)$. In this setup, we set $\Gamma=1$.
For the fourth setup, a far-neighbor term corresponding to the range $d=10$ is chosen, again with nonnegative uniform random couplings.
The fifth (and last) setup is analogous to the previous one, but with possibly negative (nearest neighbor and far-neighbor) couplings  sampled from  $\mathcal{P}(J)$.
Note that this choice leads to frustration effects, which are known to favor emergence of competing states. Therefore, we expect this setup to be particularly challenging for supervised learning algorithms.
In the presence of beyond nearest-neighbor interactions, the ground-state energy cannot be determined via the Jordan-Wigner transformation.
We resort to exact numerical diagonalization of the Hamiltonian matrix represented in the basis of eigenstates of the  $\sigma^z_j$ Pauli matrices. The Hamiltonian is stored in computer memory via a sparse representation, and the diagonalization is performed with the MKL Intel library.
Due to the exponential increase of the size of the Hilbert space with the number of spins, the accessible system sizes are  limited to $N_s=25$. 
To generate a copious enough test set for larger Ising chains, we employ a recently introduced self-learning projective quantum Monte Carlo algorithm~\cite{pilati2019self}.
This algorithm provides  unbiased stochastic estimates of the ground-state energy~\cite{pilati2020simulating},  affected only by statistical fluctuations of the order of $10^{-4}\%$. As shown in  Ref.~\cite{pilati2019supervised}, such small random fluctuations do not affect the training process.
The largest chain size we address is $N_s=50$, which is far out of reach for exact diagonalization calculations. 
These data represent a challenging benchmark for the accuracy of deep neural networks for quantum systems.
It is worth mentioning that artificial neural networks have been recently employed also to accelerate the search  of the ground-state of classical disordered  Ising models~\cite{mcnaughton2020boosting}.

\section{Artificial neural networks}
\label{secnetworks}
This Section describes the networks considered in this article.
The first network is a conventional dense feed-forward architecture. Its structure is visualized in Fig.~\ref{fig2}.
The  input layer includes $N_d$ neurons, which assume the values of the descriptors associated to each system instance. 
The  neuron $h=1,\dots,N_n^l$ in the hidden layer $l=1,\dots,N_l+1$ assumes the activation value $a_h^l=g(\sum_j w_{h,j}^l a_{j}^{l-1} + b_h^l)$. The index $j$ labels the neurons in the previous layer. The weights $w_{h,j}^l$ couple the layers $l$ and  $l-1$. With $b_h^l$ we indicate the bias terms.
For the activation function $g(x)$, we choose the rectified linear unit $ g_{\mathrm{relu}}(x) = \mathrm{max}(0,x)$, or the exponential linear activation $g_{\mathrm{elu}}(x) = x$ if $x > 0$, and $g_{\mathrm{elu}}(x) = \exp(x)-1$ if $ x \le 0$, as specified in the Appendix~\ref{Appendix}.
The output layer $l=N_l+1$ includes one neuron only. Its activation function is the identity. 
It is worth emphasizing that in the dense architecture all neurons are coupled to all neurons in the adjacent layers with distinct weights and biases.

The training process consists in optimizing the coefficients $w_{h,j}^l$ and $b_h^l$ so that when the network is fed with the $N_d$ dimensional descriptor vector $\bold{f}_t$ of a system instance, the activation of the output neuron, indicated in the following as $F(\bold{f}_t)$, closely approximates the target value $y_t$. 
The index $t=1,\dots,N_t$ labels the instances in the training set.
The optimization algorithm minimizes the loss function $\mathcal{L}(\bold{W})=\frac{1}{N_t}\sum_t \left(F(\bold{f}_t)-y_t\right)^2 +  \alpha \| \bold{W}\|_{2}^2$, including the mean-squared error and a regularization term based on the $\ell_2$-norm $\| \bold{W}\|_{2}$ of the vector formed with the weights. 
For the networks and the datasets considered in this article, the regularization parameter $\alpha$ is tuned to negligible values, unless otherwise specified.
%
The optimization method  we adopt is the Adam algorithm~\cite{kingma2014adam}.

The dense networks are not scalable, since the number of weights and biases depends on the number of descriptors. This implies that this architecture can operate only on a unique input size.
Furthermore, using dense networks for large systems is impractical. In fact,  to be flexible enough to accurately approximate general functions, the required number of hidden neurons $N_n$ has to be of the order of $N_d$. 
As discussed in Section~\ref{secresults}, $N_d$ is proportional to the system size ($L$ or $N_s$, depending on the physical system under consideration).
This implies an approximately quadratic growth of the number of weights with the system size. As a consequence, the training process requires rapidly divergent training-set sizes and computational times.
This problem is often encountered when applying artificial neural networks to image analysis. 
%
For this reason, images are usually analyzed via convolutional networks.
The structure of one of the convolutional networks employed in this article is visualized in Fig.~\ref{fig3}.
These networks include $N_c$  proper convolutional layers, followed by $N_l$  dense layers with all-to-all inter-layer connectivity. In standard practice, the last convolutional layer is connected to the dense part through the so-called flatten layer, whereby all activations are concatenated in a unique one-dimensional array. 
Notice that in Fig.~\ref{fig3} the standard flatten layer is replaced by a global pooling layer, as explained in the next paragraph.
Each convolutional layer includes a certain number of filters $N_f$.  These filters  create filtered maps of the previous layer via a convolution operation. The filtered maps are associated to the activations of $N_f$ parallel hidden layers, organized in depth within each convolutional layer.
The neuron activations are again computed via an activation function evaluated on a linear combination of the activations of the previous layer, similarly as in dense networks. However, in the convolutional layers each neuron is connected only to a tile of size $f$ of the previous layer, covering the full depth (with different weights). Theses tiles (in general) partially overlap and they span the whole layer. We consider unit stride and zero padding, so that the size of the filtered maps coincides with the input size $N_d$. Notably, all neurons in a hidden layer are coupled to their corresponding tile with the same weights and biases. This drastically reduces the number of parameters to be optimized compared to a dense network, in particular for large systems. 
%
%
Another frequently adopted strategy to boost the training process consists in inserting so-called local pooling layers between some or all convolutional layers. The pooling layers down-sample the neurons by performing
averages, or by selecting the maximum value, of the activations within small tiles  of size $p$. 

The remarkable efficiency of the convolutional networks for large systems originates from the ability of the convolutional layers to automatically  extract the most relevant features out of the (possibly many) system descriptors. Notably, this avoids recurse to hand-crafted features, making the networks quite generally applicable. 
The role of the dense layers is to operate high-level operations on the extracted features. 
However, a standard convolutional network as the one described above is not scalable. In fact, while the convolutional layers are scalable by construction, the size of the flatten layer scales with the input system size. This implies that at least the final dense layers  have to be retrained for each system size.
In this article, we implement a fully scalable convolutional network by introducing a global pooling layer in the place of the flatten layer.
We consider pooling with  the maximum or the average operation, as specified in Section~\ref{secresults}.
The activations in the global pooling layer correspond to the average or to the maximum values of the whole filtered maps in the last convolutional layer. 
They differ from more conventional local pooling layers, which perform pooling operations on small tiles.
Clearly, the number of neurons in the global pooling layer coincides with the number of filters in the last convolutional layer, and it is therefore independent on the system size. This makes the network fully scalable.
We demonstrate in Section~\ref{secresults} that convolutional networks with a global pooling layer, which we refer to as global (maximum or average) networks, display stable performances for increasing system sizes. Furthermore, the scalability property allows one to train them via transfer learning from small to large systems, and to perform extrapolations to sizes larger than the ones included in the training set.

Recently, another strategy to implement a scalable network has been introduced~\cite{mills2019extensive}. A caricature of the corresponding structure is shown in Fig.~\ref{fig4}. This strategy is designed for the extensive properties  of solid-state systems. It consists in dividing the system into a set of partially overlapping tiles.
Each tile is formed by a focus region including $\mathcal{F}$ descriptors, and by a context region including $\mathcal{C}$ descriptors. The context regions overlap with the focus regions of the adjacent tiles. 
In our implementation, the tile subdivision satisfies the periodic boundary conditions.
The descriptors corresponding to each tile are fed to a conventional convolutional network. 
The output value of each network represents the contribution of the corresponding tile to the total ground-state energy, which is computed as the sum of these contributions. The role of the context regions is to take into account the boundary effects. An important feature is that the parallel convolutional networks corresponding to each tile share the same coefficients. 

The networks described above, namely the dense,  the global (maximum and average), and the extensive networks, are implemented using the Keras library for machine learning~\cite{chollet2015keras}, with TensorFlow backend.
The structural parameters of all models, including number and type of layers and of neurons, are detailed in the Appendix~\ref{Appendix}.

\begin{figure}[h]
\begin{center}
\includegraphics[width=1.0\columnwidth]{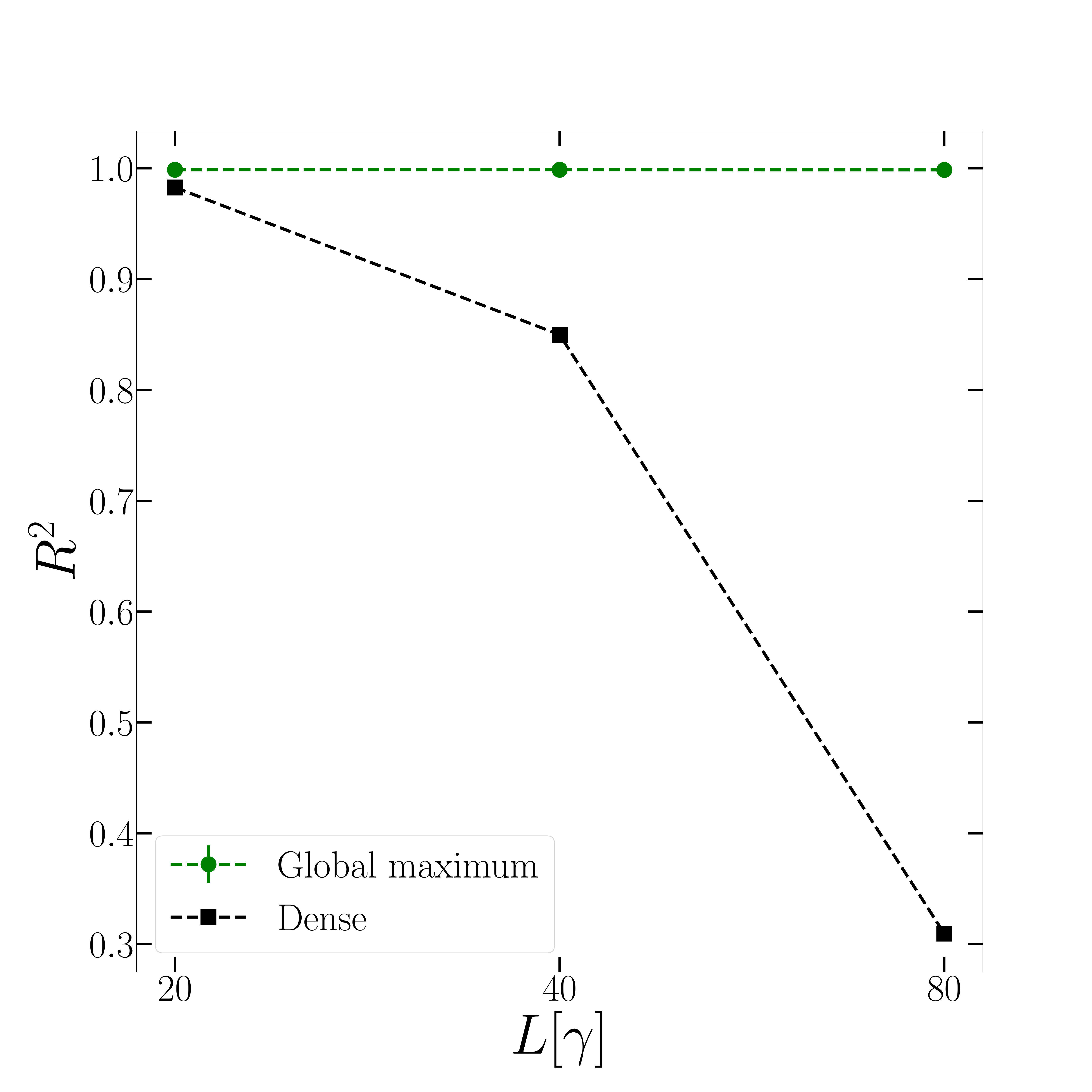}
\caption{(color online). 
Coefficient of determination $R^2$ on the test set for the one-dimensional continuous-space Hamiltonian~\eqref{H}.
Two architectures are considered, namely the global maximum and the dense networks.
They are trained and tested on the system size $L$ (horizontal axis).
Here and in the following figures, the error-bars represent the estimated standard deviation of the average over several repetitions of the training process with different pseudo-random numbers. The connecting segments are guides to the eye.
}
\label{fig5}
\end{center}
\end{figure}
%

\begin{figure}[h]
\begin{center}
\includegraphics[width=1.0\columnwidth]{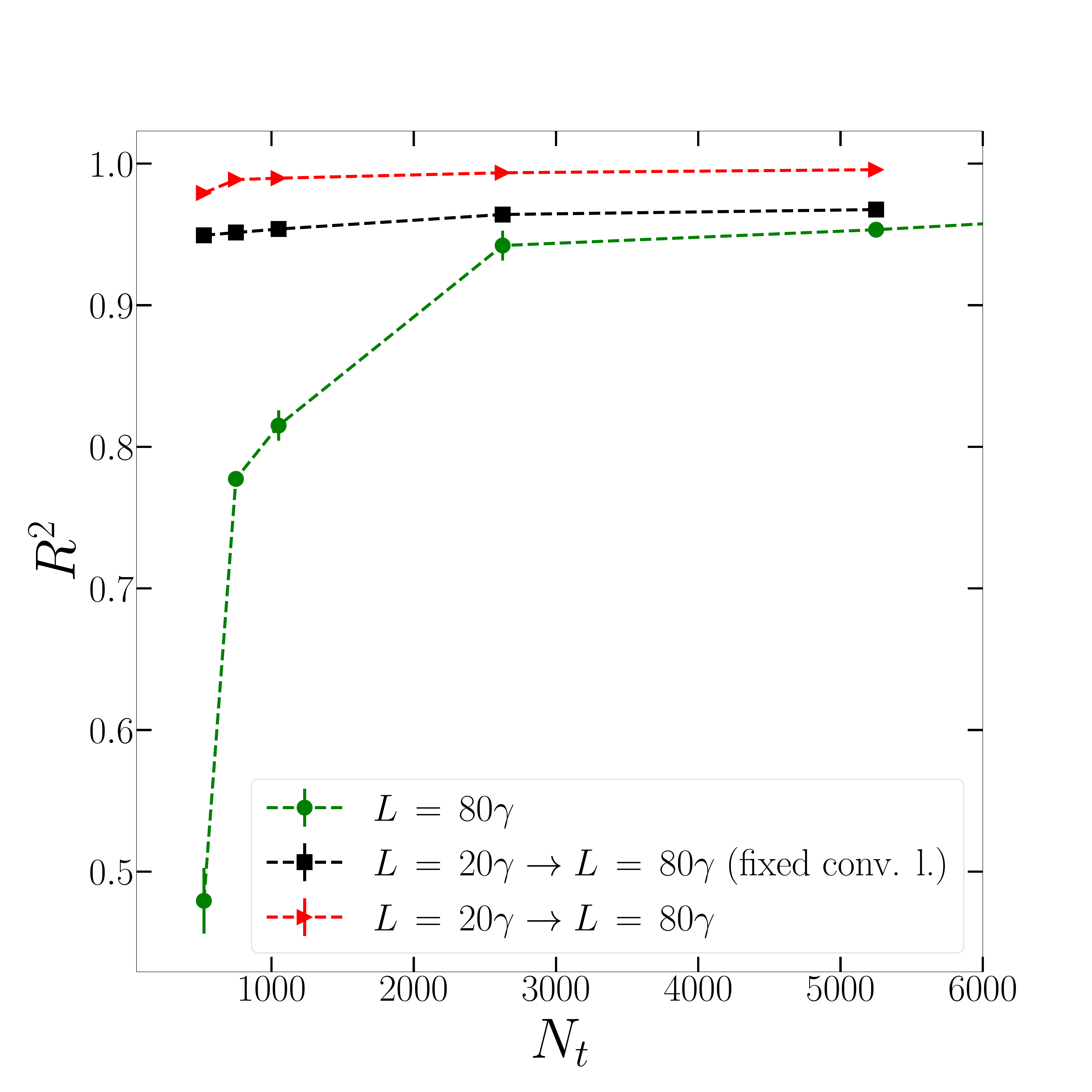}
\caption{(color online). 
Transfer learning for the one-dimensional continuous-space Hamiltonian~\eqref{H}.
The coefficient of determination $R^2$ for the test  size $L=80\gamma$ is shown as a function of the number of instances in the training (or re-training) set $N_t$.
Three global maximum networks are compared: the first one is trained from scratch on the size $L=80\gamma$  without pre-training (green circles); the second one is  pre-trained on $L=20\gamma$ (red triangles);  the third is  pre-trained on $L=20\gamma$ and only its dense layers are retrained on $L=80\gamma$ (black squares).
}
\label{fig6}
\end{center}
\end{figure}
%

\begin{figure}[h]
\begin{center}
\includegraphics[width=1.0\columnwidth]{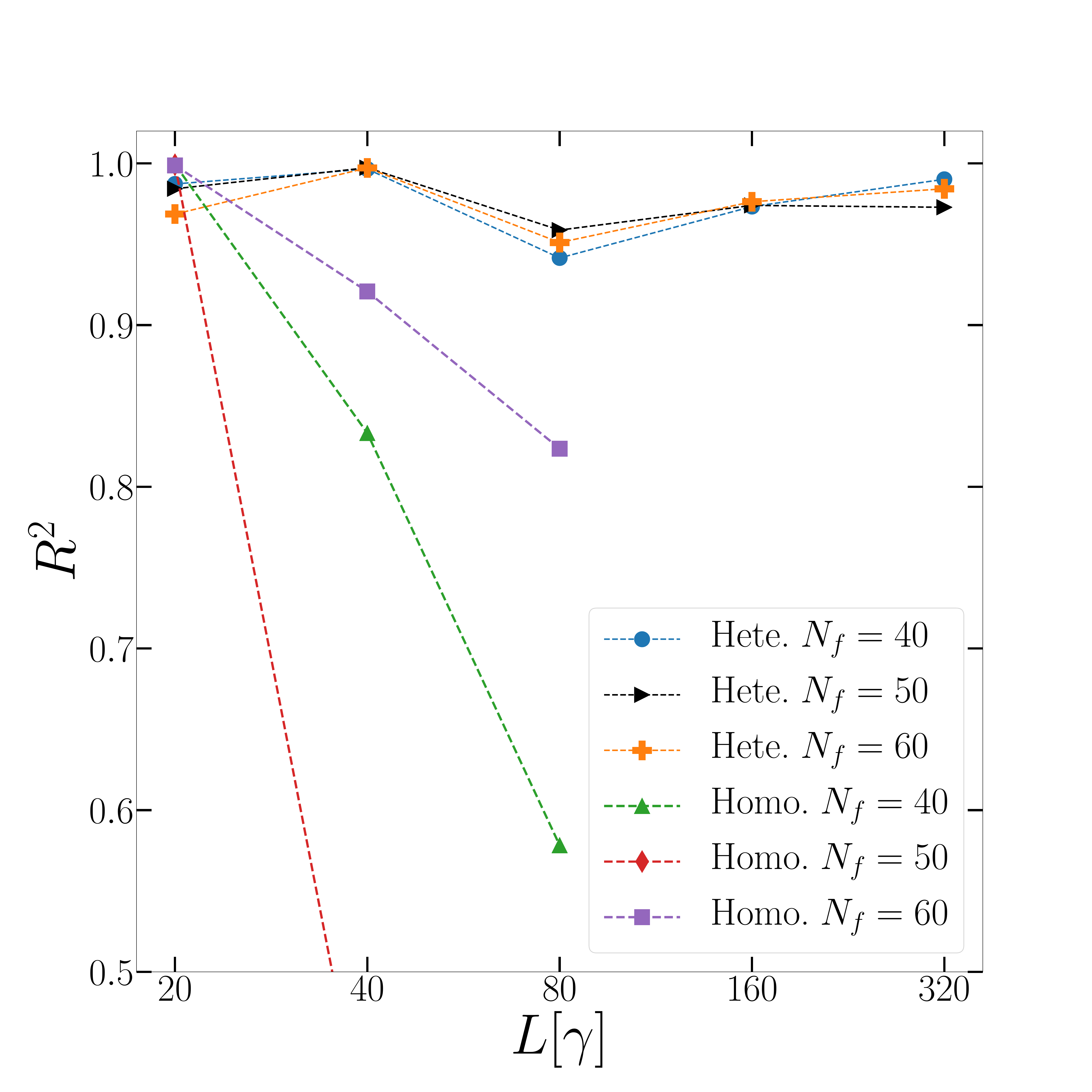}
\caption{(color online). 
Extrapolation for the one-dimensional continuous-space Hamiltonian~\eqref{H}.
The coefficient of determination $R^2$ is shown as a function of the size $L$ of the test systems.
Three global maximum networks with different numbers of convolutional filters $N_f$ are compared.
The first three symbols in the key correspond to homogeneous training on the systems size $L=20\gamma$. 
The last three symbols correspond to heterogeneous training on the systems sizes $L=20\gamma$ and $L=40\gamma$. 
}
\label{fig7}
\end{center}
\end{figure}
%

\section{Results}
\label{secresults}
The networks described in Section~\ref{secnetworks} are trained to predict the ground-state energies of the disordered Hamiltonians defined in Section~\ref{secmodels}. 
The first testbed we consider is the single-particle continuous-space model~\eqref{H}. The system instances are represented by spatial descriptors corresponding to the values of the disordered potential $V(x)$ on a discrete grid with fixed spacing $\delta x$ (see Section~\ref{secmodels}). Notice that for the smallest size we consider, namely $L=20\gamma$, the number of descriptors is as large as $N_d=1024$, and it increases proportionally to $L$.
To quantify the network's accuracy, we compute the coefficient of determination, defined as:
\begin{equation} 
 R^2=1- \frac{ \sum_{k=1}^{N_k} \left(F(\bold{f}_k)-y_k\right)^2} {  \sum_{k=1}^{N_k} \left(y_k-\bar{y}\right)^2}. 
 \end{equation}
Here, $y_k$ corresponds to the ground-state energy of the test instance $k$, and $N_k$ is the number of instances in the test set. $\bar{y}$ is the average ground-state energy of the test set.
 Clearly, none of the instances of the test set is included in the training set.
If the predictions are exact one obtains $R^2=1$. The constant function $F(\bold{f})=\bar{y}$ corresponds to the score $R^2=0$. 
$R^2$ is a fair figure of merit since it takes into account the different intrinsic variances of the target values in the different benchmarks.
In particular, these variances change with the system size.
The performances of the different networks are compared in Fig.~\ref{fig5}, as a function of the system size $L$.
For all benchmarks we consider, the training set includes $N_t \approx 10^5$ instances, while the test set includes  $N_k\approx 10^3$ or $N_k\approx 10^4$ instances, unless otherwise specified.
One notices that the accuracy of the dense network substantially decreases with $L$. We attribute this loss of performance to the rapid increase of the number of coefficients to be optimized (see Section~\ref{secnetworks}). Optimizing them would require even larger training sets, which are computationally prohibitive, and/or more powerful optimization algorithms.
The global network retains excellent performance $R^2\simeq 1$ with the available training sets. Here, the global pooling layer extracts the maximum value. This choice is motivated by the idea that in an Anderson localized system, as the one under investigation, the network has to identify the spatial region where localization occurs.
The extensive network displays unremarkable performance (data not visualized). For example, for $L=80\gamma$ the highest score we obtain is $R^2 \simeq 0.70$. This is not surprising, since in the single-particle model the ground-state energy is not extensive. In fact, since deep wells are more likely to occur in larger optical speckle fields,  low energy values become more likely  as $L$ increases. 
Still, it is worth pointing out that the extensive network outperforms the dense model.
The scalability property allows one to adopt the  transfer-learning protocol commonly employed by computer scientists working on image analysis. In that context, transfer learning is implemented by pre-training very deep networks  on large databases of generic  images from the world-wide-web, comprising $\sim 10^6$ images. Then, the networks are specialized on the desired classification task on the available, typically relatively small, training set.
Relevant examples of pre-trained deep neural networks for image analysis are the ResNet~\cite{he2016deep} and the VGG models~\cite{simonyan2014very}.
Transfer learning has recently proven very helpful also in quantum-physics research~\cite{zen2020transfer}. It has been employed to accelerate the optimization -- performed via a reinforcement learning algorithm -- of variational wave-functions built with a (non-scalable) generative neural network, namely, the restricted Boltzmann machine. 
In this article, a transfer learning protocol is used to accelerate the supervised learning of large-system properties via a pre-training performed on smaller systems, for which copious training sets can be created at a limited computational cost.
Our protocol exploits the scalability property, meaning that a network with the same structure (number of neurons and connectivity) can address different system sizes.
As shown in Fig.~\ref{fig6}, the global maximum network, when pre-trained on the system size $L=20\gamma$, reaches the remarkable performance $R^2\simeq 0.98$  for the size $L=80\gamma$ with as few as $N_t \approx 5\times 10^2$ instances in the retraining set. For larger $N_t$, the score rapidly converges towards the ideal performance $R^2 = 1$.
Instead, extremely small sets with $N_t \ll 5\times 10^2$ instances for the retraining stage  become problematic, due to the risk of overfitting the (few) available instances. 
An equivalent  network trained from scratch (without pre-training) on the $L=80\gamma$ systems requires at least as many as $N_t \sim 5\times 10^4$ instances to reach $R^2\simeq 1$.
It is also worth mentioning that, for small $N_t\lesssim 10^3$, the $R^2$ scores obtained without pre-training require appropriate tuning of the regularization parameter.
Fig.~\ref{fig6} displays also the results of a partial optimization in the retraining stage. In this protocol, the coefficients of the convolutional layers are fixed at the values obtained in the pre-training stage. This procedure is potentially useful since it significantly reduces the computational cost of retraining.
Again, the learning of the larger system is significantly accelerated compared to the optimization without pre-training. This indicates that once the convolutional layers have learned how to extract the relevant features, they can be transferred to different sizes.
Clearly, the fully retrained network reaches higher accuracy due to the superior flexibility.
The remarkable efficiency of the transfer learning protocol motivates us to attempt performing  extrapolations to system sizes larger than those included in the training set. The results are reported in Fig.~\ref{fig7}. Three global maximum networks with different numbers of convolutional filters $N_f$ are considered.
These networks are trained on a heterogeneous set, including as many as $1.6\times 10^5$ instances of size $L=20\gamma$, and also a much smaller set of $4.5\times 10^3$ instances of size $L=40\gamma$. The training process includes small cycles of $5$ epochs for each system size in an alternated fashion. 
Remarkably, these networks provide accurate predictions up to the largest size considered in our tests, namely $L=320\gamma$.
The heterogeneous training with two system sizes is essential. It ensures that the network does not specialize on a unique size. In fact, three analogous networks trained on a homogeneous set with only $L=20\gamma$ instances fail to accurately extrapolate to larger systems (see Fig.~\ref{fig7}).
%

\begin{figure}[h]
\begin{center}
\includegraphics[width=1.0\columnwidth]{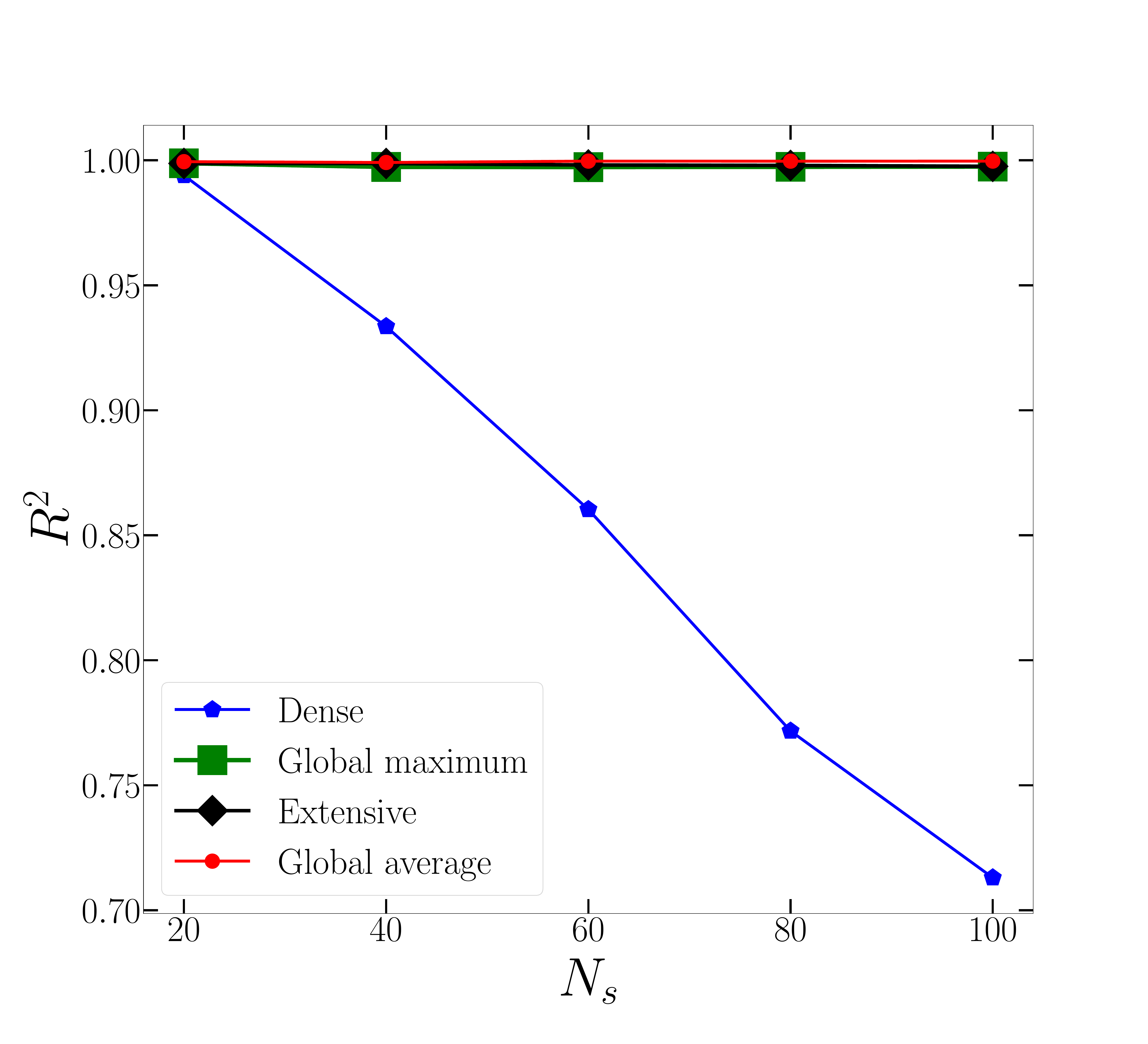}
\caption{(color online). 
Coefficient of determination $R^2$ on the test set for the quantum Ising chain~\eqref{H} with only nearest-neighbor couplings (first setup, see Section~\ref{secmodels}).
Four networks are considered: the dense, the global maximum, the extensive, and the global average networks.
They are trained and tested on the system sizes $N_s$ (horizontal axis).
}
\label{fig8}
\end{center}
\end{figure}
%

\begin{figure}[h]
\begin{center}
\includegraphics[width=1.0\columnwidth]{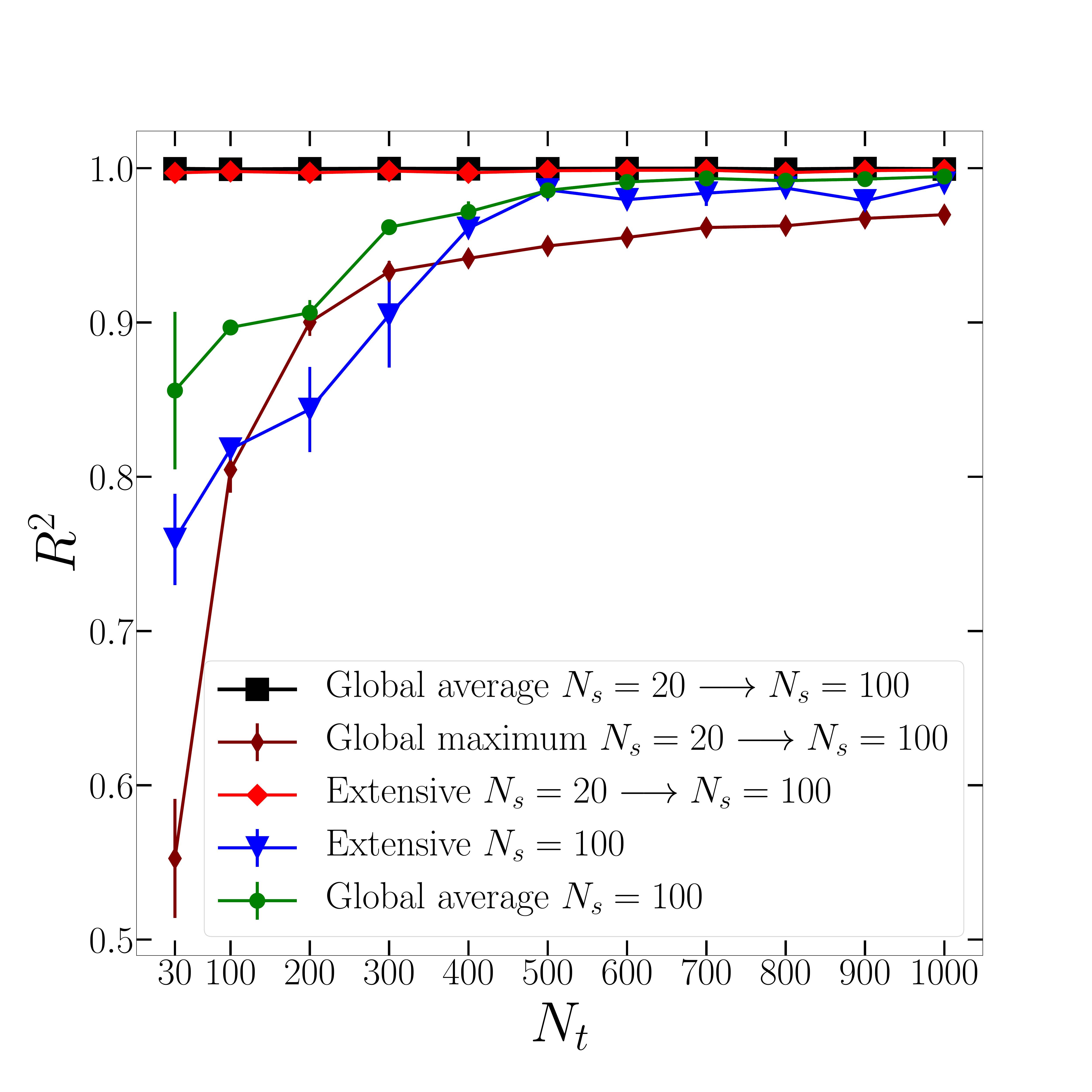}
\caption{(color online). 
Transfer learning for the quantum Ising chain~\eqref{H} with only nearest-neighbor interactions (first setup).
The coefficient of determination $R^2$ for the test system size $N_s=100$ is shown as a function of the number of instances in the training (or retraining) set $N_t$.
Three networks are compared: the global average, the global maximum, and the extensive networks.
The first three symbols in the key correspond to networks pre-trained on the size $N_s=20$, and then retrained on $N_s=100$. The last two symbols correspond to networks obtained without pre-training.
}
\label{fig9}
\end{center}
\end{figure}
%

\begin{figure}[h]
\begin{center}
\includegraphics[width=1.0\columnwidth]{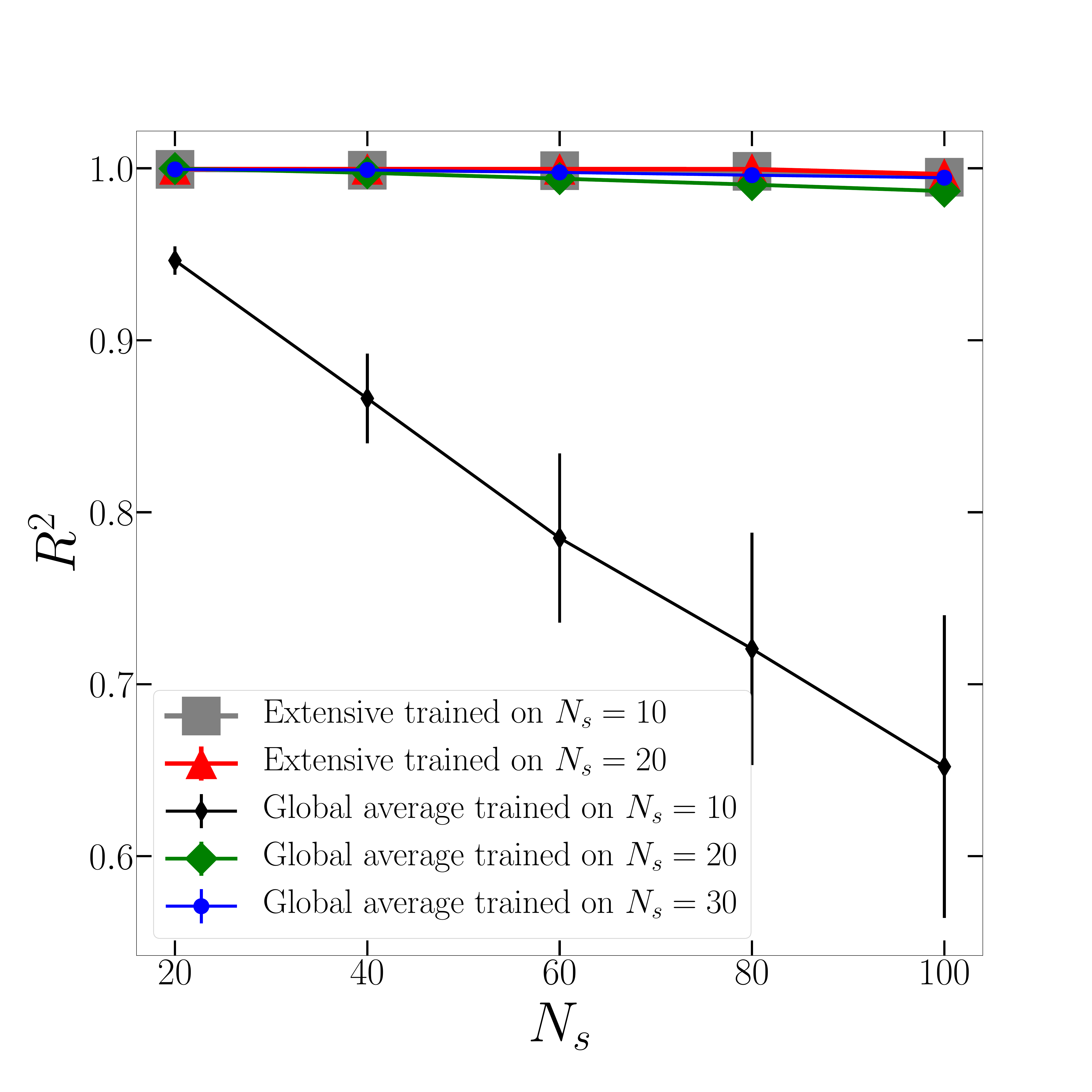}
\caption{(color online). 
Extrapolation for the quantum Ising chain~\eqref{H} with only nearest-neighbor interactions, tuned at the quantum critical point (second setup).
The coefficient of determination $R^2$ is shown as a function of the test system size $N_s$.
The global average network and the extensive network are compared. The training is performed on the sizes indicated in the key. 
}
\label{fig10}
\end{center}
\end{figure}
%

\begin{figure}[h]
\begin{center}
\includegraphics[width=1.0\columnwidth]{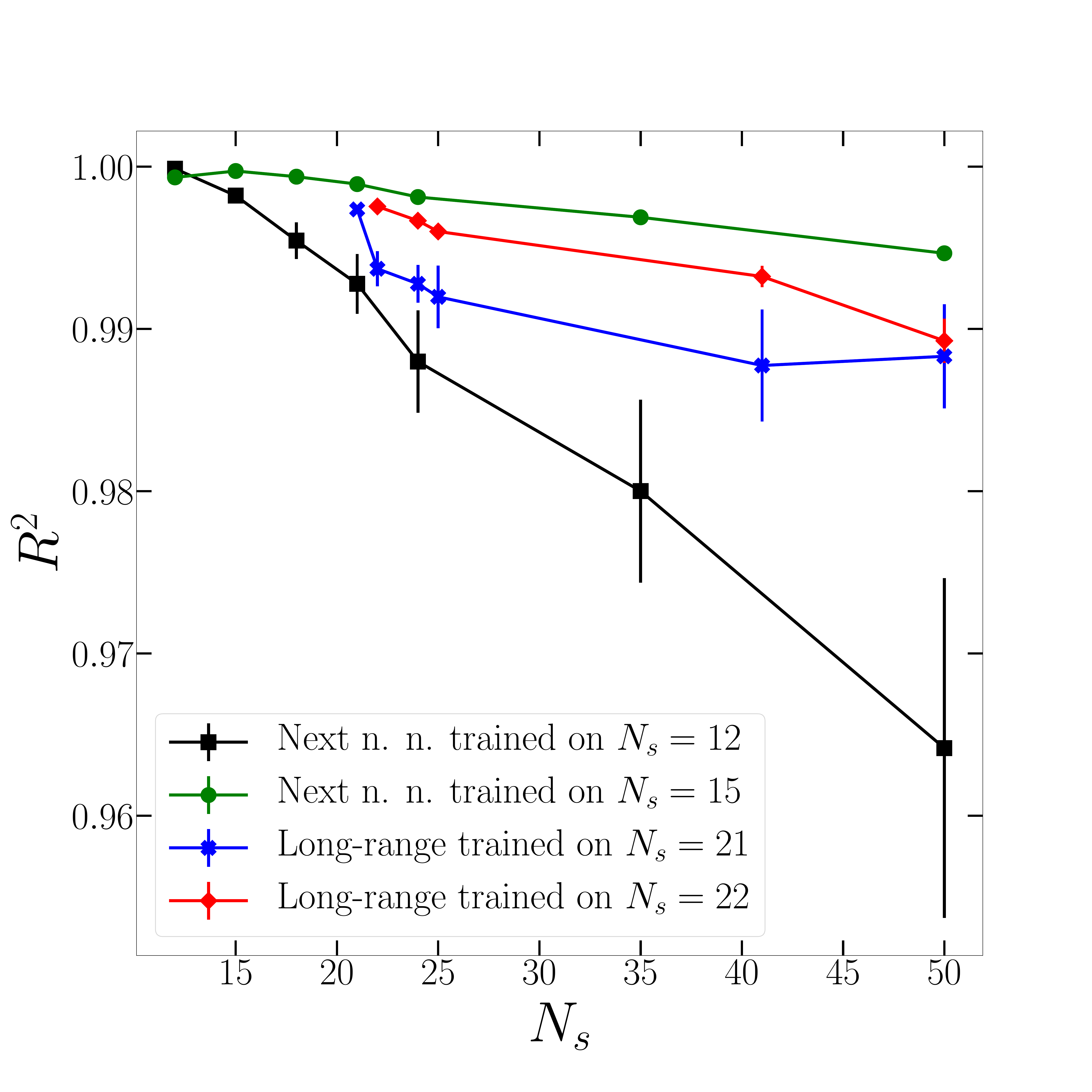}
\caption{(color online). 
Extrapolation for the quantum Ising chain~\eqref{H} in two setups  (third and fourth setups, see text): nearest neighbor plus  next-nearest-neighbor interactions (i.e., range $d=2$); nearest neighbor plus long-range interaction (range $d=10$) .
The coefficient of determination $R^2$ is shown as a function of the test system size $N_s$.
The global average network is considered, and it is trained on the sizes indicated in the key. 
}
\label{fig11}
\end{center}
\end{figure}
%

The second testbed we consider is the random quantum Ising chain~\eqref{H2}. Here, the ground-state energy $E$ is extensive, since on average it increases with the system size $N_s$. 
For the first (and also the second) setup (see details in Section~\ref{secmodels}), which include only nearest-neighbor interactions, $E$ is efficiently obtained via the Jordan-Wigner transformation.
The $N_s$ nearest-neighbor couplings $J_j$ are used as system descriptors.
Fig.~\ref{fig8} analyses the performances of different networks, training and testing on the same system size. 
The training set includes $N_t\simeq 3 \times 10^4$ instances.
While the accuracy of the dense network rapidly decreases as the system size increases, the extensive network and the two types of global networks maintain $R^2 \simeq 1$. 
For the extensive network, the optimal focus and context sizes are $\mathcal{F}=1$ and $\mathcal{C}=1$, respectively, as previous found in a study on classical spin models~\cite{mills2019extensive}. 
In the first  global network, the global pooling layer extracts the maximum values of the filtered maps. In the second, it computes the average value. One might expect the latter choice to be more appropriate for the extensive property under consideration. This expectation is confirmed below.
Transfer learning is analyzed in Fig.~\ref{fig9}.  The global average network and the extensive network, when pre-trained on the small size $N_s=20$, accurately predict the ground-state energy of $N_s=100$ spin chains even with as few as $N_t=30$ instances in the retraining stage. To reach a comparable accuracy without pre-training, they need at least $N_t \approx 10^3$ instances. As anticipated, the global maximum network is much less efficient, requiring $N_t \gg 10^3$ even with pre-training.
The accuracy of the extrapolations is analyzed in Fig.~\ref{fig10}. Here, we consider the second setup. It corresponds to the quantum critical point of the Hamiltonian~\eqref{H2} with only nearest-neighbor interactions. The critical point is particularly interesting since long-range (ferromagnetic) correlations develop. In fact, one might speculate these correlations to be difficult to describe by neural network models. In Fig.~\ref{fig10}, a comparison is made between networks trained on different system sizes. Both the global average network and the extensive network demonstrate remarkable accuracies, even for the largest system sizes considered in our tests. For the former network, the performance improves if pre-training is performed on larger sizes. For the latter, the performance is remarkable even when training is performed on the smallest system size $N_s=10$.
The next benchmarks we consider are the third and fourth setups of the random Ising chain (see Section~\ref{secmodels}). Beyond the nearest-neighbor interaction, they include a next-nearest-neighbor term (corresponding to the range $d=2$) and a far-neighbor term ($d=10$), respectively. 
The uniform random couplings $J_i$ and $J^{\prime}_j$ are non-negative. This choice avoids frustration effects.
These Hamiltonians represent a more challenging benchmark, since one cannot speculate that the networks are simply learning to perform the Jordan-Wigner calculation. It is also worth mentioning that long-range interactions, e.g., electrostatic forces, are not addressed by common supervised-learning models for electronic systems~\cite{bartok2017machine}. We argue that the $d=10$ term considered here is suitable to test whether the networks can capture non-local effects.
In these two setups, the system instances are represented by a descriptor vector of size $N_d=2N_s$, including the nearest-neighbor and the beyond nearest-neighbor couplings
in the alternate ordering $\bold{f}=\left(J_1,J^\prime_1,J_2,J^\prime_2,\dots, J_{N_s},J^\prime_{N_s}\right)$.
In Fig.~\ref{fig11} we analyze the extrapolations of two global average networks, trained on the sizes $N_s=12$ and $N_s=15$ for the $d=2$ setup, and on $N_s=21$ and $N_s=22$ for the  $d=10$ setup.
Remarkably, these networks accurately reproduce the ground-state energies of $N_s = 50$ spin chains, which are computed via quantum Monte Carlo simulations~\cite{pilati2020simulating} ($N_k \simeq 200$ test instances are considered). The performance improves when training is performed on the larger size.
The last benchmark we consider is a $N_s=21$ spin chain in the fifth setup, which includes nearest-neighbor and far-neighbor interactions ($d=10$) with both positive and negative (random) couplings, leading to frustration effects. 
The representation we adopt in this case is based on a two-channel descriptor matrix of size $N_s \times 2$, where each row includes the pair ($J_j$, $J_j^\prime$), for $j=1,\dots,N_s$.
The global average network provides accurate predictions, namely $R^2  = 0.991(3)$, even for this challenging 
benchmark~\footnote{With a single channel description, we obtain $R^2=0.980(5)$. 
Furthermore, a preliminary analysis  with the extensive network provides mediocre results, possibly indicating a difficulty in describing long-range interactions}.
These findings indicate that the scalable networks provide accurate predictions  for large-scale   complex quantum systems.
Notably, these predictions are obtained at a minuscule computational cost, giving access to system sizes which are out of reach for  other accurate computational techniques.

\section{Conclusions}
\label{secconclusions}
We have analyzed the scalability of different artificial neural networks  in the supervised learning of disordered quantum systems. 
The accuracies of the networks have been rigorously quantified as a function of the system size.
We have introduced a scalable network based on a convolutional architecture complemented by a global pooling layer. This layer allows the network addressing arbitrary system sizes.
This scalable architecture has been compared to (non-scalable)  dense networks and to the (scalable) extensive network introduced in Ref.~\cite{mills2019extensive}. 
These networks have been tested for various  benchmarks, namely a  continuous-space single-particle model relevant for cold-atom experiments, and a random quantum Ising chains in different setups, including one with only short-range interaction, one with long-range and frustrated interactions, and one tuned at the ferromagnetic quantum critical point.
The novel scalable architecture retains high accuracy as the system size increases, both for the (intensive) continuous-space Hamiltonian and for the (extensive) quantum Ising chain. As expected, the extensive network performs well only in the latter testbed. Both networks outperform conventional dense networks with all-to-all connectivity.
As discussed in Section~\ref{secresults}, the scalability property allows one to adopt the transfer-learning protocol familiar from the field  of image analysis. As we demonstrated, this protocol allows accurately learning the properties of large quantum systems with small training sets, orders of magnitude sparser than those required without pre-training. 
Remarkably, the scalable architecture is also able to extrapolate to system sizes larger than those included in the training set, accurately reproducing the results of state-of-the-art quantum Monte Carlo simulations.

Our study highlights the crucial role of the network scalability in the supervised learning  of complex quantum systems. 
This property allows accurately learning large-system properties with computationally feasible training sets. It also gives access to sizes that are out of reach for other accurate computational techniques.
We demonstrated that scalable networks can learn exact ground-state properties, not only approximate results based on, e.g., density functional theory, as shown in previous studies. Notably, our testbeds included long-range interactions, which have been often omitted in previous investigations.
The scalable network we have introduced does not require manual development of ad-hoc features. In fact, it can be applied to rather generic physical systems with straightforward system representations, allowing the number of descriptors to scale with the system size.
For example, this network  could be directly applied to solid-state or molecular systems, predicting properties such as atomization or ionization energies. As descriptor vectors, one could employ images representing the atomic density distribution, as in Ref.~\cite{mills2019extensive}. 
The instances in the training set and in the test set would represent different molecular configurations or crystals with different porosities obtained by randomly removing a certain number of atoms.
We leave these endeavors to future investigations.

\section*{Acknowledgements}
\noindent
We  acknowledge useful discussions with P. Mujal, A. Polls, B. Juli\'{a}-D\'{i}az, P. Pieri, and A. Perali.\\
S. P.  and S. C. acknowledge financial support from the FAR2018 project titled ``Supervised machine learning for quantum
matter and computational docking'' of the University of Camerino and from the Italian MIUR under the project PRIN2017 CEnTraL 20172H2SC4.
S. P. also acknowledges the CINECA award under the ISCRA initiative, for the availability of high performance computing resources and support.

\appendix
\section{Details of the artificial neural networks}
\label{Appendix}
Here we provide further details of the networks we employed to obtain the results presented in the figures indicated below:

\begin{itemize}

\item Figs.~\ref{fig5} to ~\ref{fig7}: the dense network has $N_l=5$ hidden layers with $N_n=100$ neurons. The activation function is $g_{\mathrm{relu}}(x)$. As in all models discussed below, there is an additional output layer including a single neuron with either $g_{\mathrm{relu}}(x)$ activation function (for the continuous-space model) or identity activation function (for the spin models).\\
The global maximum network has $N_c=6$ convolutional layers with $N_f=50$ filters of size $f=5$, separated by local maximum pooling layers with tile size $p=3$. The global maximum pooling layer is followed by $N_l=2$ dense layers with $N_n=10$ neurons. The activation function is $g_{\mathrm{relu}}(x)$
\item Figs.~\ref{fig8} to~\ref{fig11}: 
The dense network has $N_l=4$ hidden layers with $N_n=100$. The activation function is $g_{\mathrm{elu}}(x)$.\\
The global maximum network has $N_c=5$ convolutional layers with $N_f=30$ and $f=3$. They are separated by local maximum pooling layers with  $p=2$. The global maximum pooling layer is followed by $N_l=2$ dense layers with $N_n=10$. The activation function is $g_{\mathrm{relu}}(x)$. \\
The global average network has two sets of $N_c=3$ convolutional layers with $N_f=30$ and $f=3$. The two sets are separated by a local average pooling layer with  $p=2$. The global average pooling layer is followed by $N_l=2$ dense layers with $30$ and $20$ neurons, respectively. The activation function is $g_{\mathrm{elu}}(x)$. \\
The extensive network is formed by identical parallel convolutional networks with focus size $\mathcal{F}=1$ and context size $\mathcal{C}=1$. 
Two sets of $N_c=2$ convolutional layers with $N_f=30$ and $f=3$ are followed by a local maximum pooling layer with $p=2$. The flatten layer is followed by $N_l=2$ dense layers with $N_n=10$, and by another (single neuron) output layer (with identity activation function). The output values are summed.  The activation function is $g_{\mathrm{elu}}(x)$.\\
For the extensive network, the  target value is $E$, while for the first three models it is $E/N_s$.

\item Random Ising chain~\eqref{H2}, fifth setup (see Section~\ref{secmodels}): the global average network has $N_c=6$ convolutional layers with $N_f=20$ and $f=4$. The global pooling layer is followed by $N_l=2$ dense layers with $N_n=20$. The activation function is $g_{\mathrm{elu}}(x)$. The target value is target value is $E/N_s$.

\end{itemize}

\bibliography{Ref}

\end{document}